\documentstyle[12pt]{article}

\def\rdots{\mathinner{\mkern1mu\raise1pt\vbox{\kern1pt\hbox{.}}\mkern2mu
  \raise4pt\hbox{.}\mkern2mu\raise7pt\hbox{.}\mkern1mu}}
\newcommand{\be}{\begin{equation}}
\newcommand{\ee}{\end{equation}}
\newcommand{\Z}{{\rm Z\kern-.35em Z}}
\newcommand{\bP}{{\rm Z\kern-.35em Z}}
\newcommand{\Q}{\kern.3em\rule{.07em}{.65em}\kern-.3em{\rm Q}}
\newcommand{\R}{{\rm I\kern-.15em R}}
\newcommand{\h}{{\rm I\kern-.15em H}}
\newcommand{\C}{\kern.3em\rule{.07em}{.65em}\kern-.3em{\rm C}}
\newcommand{\T}{{\rm T\kern-.35em T}}
\newcommand{\D}{{\kern-.5em /}}

\begin{document}
\openup 1.5\jot
\centerline{Some Generalized BRS Transformations. I}
\centerline{The Pure Yang-Mills Case}

\vspace{1in}
\centerline{Paul Federbush}
\centerline{Department of Mathematics}
\centerline{University of Michigan}
\centerline{Ann Arbor, MI 48109-1109}
\centerline{(pfed@math.lsa.umich.edu)}

\vspace{1in}

\centerline{Abstract}

Some generalized BRS transformations are developed for the pure Yang-Mills theory, and a form of quantum gravity. Unlike the usual BRS transformations: these are nonlocal; may be infinite formal power series in the gauge fields; and do not leave the actio
n invariant, but only the product $e^{-S}$ with the Jacobian. Similar constructions should exist for many other field theory situations.

\vfill\eject

\noindent
I.\underline{Introduction}

Since the development of BRS transformations for the Yang-Mills theory, [1], they have played a major role in theoretical applications, such as to the study of renormalization and unitarity. BRS transformations have also been given for quantum gravity [2]
,[3],[4], and applied to study the renormalizability of higher derivative quantum gravity, [4]. Our interest was to develop a BRS transformation for a particular formulation of quantum gravity in a natural gauge to the theory, [5]. This led us to develop 
the generalized BRS transformations of this paper, and to apply them to the pure Yang-Mills theory. The Yang-Mills setting is a simpler arena to present the basic ideas, and hopefully generalized BRS transformations may have application to the Yang-Mills 
theory. There has been study of some aspects of the Yang-Mills theory by other generalizations of the BRS symmetry,[6].

For the pure Yang-Mills we write the action as follows:

\be S = \int \rm {Tr} [\alpha F_{\mu\nu}^2 +\frac {\beta} {2} (\partial_{\mu}A_{\mu})^2 + \gamma \bar {c}_i {(\partial_{\mu}^L L_i)(\partial_{\mu} L_j+ [A_{\mu},L_j])}c_j].  \ee

The superscript L indicates differentiation is to the left, and $c_i$ is the ghost field. Sum over repeated indices will always be understood, except where otherwise indicated.

The BRS transformations are then:

\be A_{\mu}(x) \rightarrow A_{\mu}(x) + (\partial _{\mu}c_j (x)) L_j \lambda + [A _{\mu}(x), L_j]c_j (x)\lambda. \ee

\be \bar{c}_j (x) \rightarrow \bar{c}_j (x) - (\frac {\beta}{\gamma}) \partial _{\mu} A_j^{\mu}(x) \lambda. \ee 

\be c_j (x) \rightarrow c_j (x) + \frac {1}{2} s_{j\kappa \ell} c_{\kappa} (x) c_{\ell}(x) \lambda. \ee 

We work in Euclidean space, and the $L_i$ are orthonormal in the trace inner product. These transformations leave the action invariant and have (super -) Jacobian 1 (of course working to linear order in $\lambda$). The structure constants satisfy:

\be s_{ij\kappa} = {\rm Tr} (L_i [L_j,L_\kappa]). \ee

II \underline{Generalized BRS Transformations for the Pure Yang-Mills}

In contrast to (2),(3),(4) the generalized BRS transformations for the pure Yang-Mills theory will involve a rather arbitrary formal gauge transformation and are given as:

\begin{eqnarray}
 A_{\mu}(x) \rightarrow A_{\mu}(x) &+&\frac {\partial}{\partial x^{\mu}}[c_j(x)+\int_{y}F_j (x,y)c_j(y)]L_j \lambda \nonumber \\
&+&[A_{\mu}(x),L_j][c_j(x)+\int_{y}F_j (x,y)c_j(y)]\lambda \end{eqnarray}

\be \bar {c}_i (x) \rightarrow \bar {c}_i(x) - (\frac {\beta}{\gamma})(\frac {\partial}{\partial x^{\mu}}A_i^{\mu}(x))\lambda - (\frac {\beta}{\gamma}) G_i (x) \lambda \ee

\be c_j(x) \rightarrow c_j (x) + \frac {1}{2} s_{j\kappa \ell} c_{\kappa}(x) c_{\ell}(x) \lambda + \int_{y} \int_{z} Z_{j\kappa \ell} (x,y,z) c_{\kappa}(y)c_{\ell}(z) \lambda \ee

Here $F_j(x,y)$ is an essentially arbitrary formal power series in the $A_{\mu}$ field with the lowest order term of degree 1.

\be F_j(x,y)=F_j^1(x,y)+F_j^2(x,y)+ ... \ee

$F_j^i (x,y)$ is of degree $i$. $F^1$, say, is of form:

\be F_j^1 (x,y) = \int_z f_j^{\mu i} (x,y,z,) A_{\mu}^i (z)\ee
where
\be A_{\mu}(x)=\Sigma_i A_{\mu}^i (x)L_i \ee

The $G_i$ and $Z_{j\kappa \ell}$ are determined as formal power series in the $A_{\mu}$, inductively by degree, as will be specified below. If $F_j \equiv 0$ one gets the usual BRS transformation. If to order one in $\lambda$ we write:

\be S \rightarrow S + \Delta S \lambda \ee
\be J=1+\Delta J \lambda \ee

Where $J$ is the       Jacobian of the transformation (6) --(8), then we require:

\be \Delta S-\Delta J=0 \ee
which ensures invariance of $\int e^{-S}$ (i.e. invariance of $e^{-S}$ times integration measure density).

We write
\be \Delta S = \Delta S_1 +\Delta S_2 \ee
\be \Delta J = \Delta J_1 +\Delta J_2 \ee
where the subscripts 1 and 2 split the expressions into terms linear and quadratic in ${c_i (x)}$. Eq. (14) becomes two equations:

\be \Delta S_1 - \Delta J_1 = 0 \ee
\be \Delta S_2 - \Delta J_2 = 0 \ee

It is easy to see:
\be \Delta J_2=0 \ee

The equations (18)--(19) are just:
\be  \Delta S_2 = 0 \ee
which by a simple calculation holds for $Z$ satisfying:

\begin{eqnarray}
\Delta _x Z_{i\kappa \ell}(x, y, z)&+& \frac {\partial}{\partial x^\mu}(A_{\mu}^{\nu}(x) Z_{s\kappa\ell}(x,y,z))s_{irs}\nonumber \\ &-& \frac{\partial}{\partial x^{\mu}}[(\frac {\partial}{\partial x^{\mu}}F_{\kappa} (x,y))\delta (z-x)]s_{i \kappa \ell}
\nonumber \\ &-& \frac {\partial}
{\partial x^{\mu}}[A_{\mu}^s (x) F_{\kappa}(x,y) \delta (z-x)]s_{rs\kappa}s_{ir\ell} =0 \end{eqnarray}
$\delta(z-x)$ is a four dimensional delta function.
Equation (21) may be solved inductively in degree for $Z$ a formal power series in the fields $A_{\mu}(x)$, similar to $F_{\kappa}(x,y)$.\\ In equation (21) indices $i, \kappa, \ell$ are never summed!

With $F_{\kappa}(x,y)$ given, and $Z_{i\kappa \ell}(x,y,z)$ now determined, $G_i(x)$ is obtained from \\ equation (17), completely specifying the generalized BRS transformation (6)--(8). Similarly to the derivation of $Z_{i\kappa \ell}(x,y,z)$, equation (
17) holds if  $G_i(x)$ satisfies;
 
\begin{eqnarray}
 \Delta_{y}G_i(y)&-&(\frac {\partial}{\partial y^{\mu}}G_r(y))A_j^{\mu}(y)s_{rji}+\int_x  A_i^{\nu}(x),_{\mu,\nu} \frac{\partial}{\partial  x^{\mu}}F_i(x,y) \nonumber \\
&+& \int_xA^\nu_{r,\mu,\nu}(x) A_j^\mu(x)F_i(x,y) s_{rji}+ \frac{1}{\beta}\int_x\frac{\delta}{\delta A^{\mu}_i(x)}\frac{\partial}{\partial x^{\mu}}F_i(x,y) \nonumber \\
&+& \frac{1}{\beta}\int_x\frac{\delta}{\delta A_r^{\mu}(x)}\left\{A_{\mu}^j(x)F_i(x,y)\right\}
s_{rji} \nonumber \\
  &-& \frac{1}{\beta}\int_x \left[ Z_{jji}(x,x,y)-Z_{jij}(x,y,x)\right]=0 .
\end{eqnarray}

This equation may be solved inductively as a formal power series in $A_{\mu}(x)$ for $G_i$.The familiar notation for functional derivative has been used, and commas indicate partial derivatives. Index $i$ is never summed over! The last term is eq.(22) is 
delicate to calculate... the nitty-gritty yet awaits a proper exegesis.

\vfill\eject

\centerline{References}
\begin{itemize}
\item[[1]] C.Becchi,A.Rouet, and R.Stora, ``Renormalization of the Abelian Higgs-Kibble Model", {\it CMP} {\bf 42} 127 (1975). C.Becchi, A.Rouet, and R.Stora, ``Renormalization of Gauge Theories", {\it Annals Phys.}{\bf 98} 287 (1976).
\item[[2]] R.Delbourgo and M.Ramon-Medrano, ``Becchi-Rouet-Stora Gauge Identities for Gravity",{\it Nucl. Phys.  B}{\bf 110}, 467 (1976).
\item[[3]] J.Dixon, unpublished Ph.D. thesis, Oxford (1975).
\item[[4]] K.S. Stelle, ``Renormalization of higher-derivative quantum gravity",{\it Phys. Rev. D,}{\bf 16} 953-969 (1977).
\item[[5]] P.Federbush, ``A Speculative Approach to Quantum Gravity",Symposium in Honor of Eyvind H. Wichmann,University of California, Berkeley, June 1999.
\item[[6]] Satish, D.Joglekar, A.Misra, ``Relating Green's Functions in Axial and Lorentz Gauges Using Finite Field-Dependent BRS Transformations", hep-th/9812101.
\item[[7]] G. Battle, P.Federbush, and P.Uhlig, ``Wavelets for Quantum Gravity and Divergence-Free Wavelets",{\it Appl. and Comp. Harmonic Analysis 1}, 295 (1994).
\end{itemize}
\end{document}